\documentclass[twocolumn,showpacs,preprintnumbers,amsmath,amssymb]{revtex4}

\usepackage{graphicx}
\usepackage{dcolumn}
\usepackage{bm}

\begin{document}

\title{Enhancement of the retrapping current of superconducting microbridges of finite length}

\author{D.Y. Vodolazov$^1$}
\email{vodolazov@ipm.sci-nnov.ru}
\author{F.M. Peeters$^2$,}
\affiliation{$^1$ Institute for Physics of Microstructures,
Russian Academy of Sciences, 603950,
Nizhny Novgorod, GSP-105, Russia \\
$^2$Departement Fysica, Universiteit Antwerpen (CGB),
Groenenborgerlaan 171, B-2020 Antwerpen, Belgium}

\date{\today}


\begin{abstract}

We theoretically find that the resistance of a superconducting
microbridge/nanowire {\it decreases} while the retrapping current
$I_r$ for the transition to the superconducting state {\it
increases} when one suppresses the magnitude of the order
parameter $|\Delta|$ in the attached superconducting leads. This
effect is a consequence of the increased energy interval for
diffusion of the 'hot' nonequilibrium quasiparticles (induced by
the oscillations of $|\Delta|$ in the center of the microbridge)
to the leads. The effect is absent in short microbridges (with
length less than the coherence length) and it is relatively weak
in long microbridges (with length larger than the inelastic
relaxation length of the nonequilibrium distribution function). A
nonmonotonous dependence of $I_r$ on the length of the microbridge
is predicted. Our results are important for the explanation of the
enhancement of the critical current and the appearance of negative
magnetoresistance observed in many recent experiments on
superconducting microbridges/nanowires.

\end{abstract}

\maketitle

\section{Introduction}

Recently, several experimental groups observed a negative
magnetoresistance (NMR) of superconducting nanowires/microbridges
at temperatures lower than the critical temperature
\cite{Xiong,Rogachev,Tian1,Tian2,Chen1,Chen2,Zgirski}. In Refs.
\cite{Tian1,Tian2,Chen1,Chen2} authors demonstrated that in their
case the effect is connected with the suppression of
superconductivity in the bulk superconductor caused by the applied
magnetic field. Moreover in Refs.
\cite{Rogachev,Tian1,Tian2,Chen1,Chen2} it was shown that this NMR
appears together with an enhancement of the critical current of
the nanowires in weak magnetic fields.

Presently there exist several theories
\cite{Pesin,Wei,Kharitonov,Fu,Vodolazov1} which propose different
mechanisms for the NMR. Ref. \cite{Pesin} claims that it is
connected with the suppression of a new channel of dissipation in
superconducting wires by an applied magnetic field, while Refs.
\cite{Wei,Kharitonov} argue that the suppression of the intrinsic
pair-breaking resulting from total 'spin-flip'+'non-spin-flip'
rate is responsible for the effect. In Ref. \cite{Fu} the
stabilization of the superconducting phase due to magnetic field
induced normal metal/superconductor boundaries at the ends of the
microbridge was offered as the main mechanism for the negative
magnetoresistance while in Refs.
\cite{Xiong,Vodolazov1,Arutyunov1} the effect was explained as due
to a decrease of the charge imbalance decay length $\lambda_Q$ in
weak magnetic fields.

Taking into account the strong dependence of both the NMR and the
enhancement of $I_c$ on the length of the superconducting
microbridge/nanowire (i.e. the effect does not exist in relatively
long and short samples - see Refs. \cite{Tian1,Tian2,Chen1,Chen2})
we argue that the proposed mechanisms as put forward in Refs.
\cite{Pesin,Wei,Kharitonov,Fu} are not relevant for these
experiments. Indeed their applicability is not limited by the
length of the superconductors. Due to the same reason a decrease
of $\lambda_Q$ \cite{Xiong,Vodolazov1,Arutyunov1} cannot explain
the length dependence of the effect because it should lead to NMR
in samples of arbitrary length. Besides at currents close to the
depairing current, $\lambda_Q$ starts to depend on the current
\cite{Stuivinga} and it smears its dependence on H. As a result
this mechanism cannot explain the increase of $I_c$ in weak
magnetic fields.

In Ref. \cite{Vodolazov1} another mechanism for the enhancement of
$I_c$ was proposed connected with the complete suppression of the
order parameter in the superconducting leads. But this mechanism
is not able to explain the monotonous enhancement of $I_c$ in weak
magnetic fields (see \cite{Rogachev,Chen1,Chen2}) when the
superconducting leads are still in the superconducting state with
weakly suppressed order parameter.

Here, we offer a new mechanism that leads to the negative
magnetoresistance of microbridges/nanowires and the enhancement of
the critical (retrapping) current. Our explanation is based on the
following idea: even a weak suppression of the order parameter in
the leads opens new energy channels for the diffusion of the 'hot'
quasiparticles from the microbridge where they are induced by
oscillations of the order parameter \cite{Vodolazov2}. As a result
the effective 'temperature' of the quasiparticles {\it in the
microbridge} decreases and the retrapping current for the
transition to the superconducting state increases. Simultaneously
it leads to decrease of the resistance of the microbridge at fixed
current. Our proposed mechanism leads to no effect in very short
and very long microbridges/nanowires (in agreement with the
experiments \cite{Tian1,Tian2,Chen1,Chen2}). Besides we find that
the retrapping current is a nonmonotonous function of the length
of the microbridge.

The paper is organized as follows. In section II we present the
theoretical model. The current-voltage (IV) characteristics of
microbridges of different length are presented in Section III for
different values of the order parameter in the leads. In section
IV we discuss our result and in Sec. V we present our conclusions.

\section{Model}

\begin{figure}[hbtp]
\includegraphics[width=0.45\textwidth]{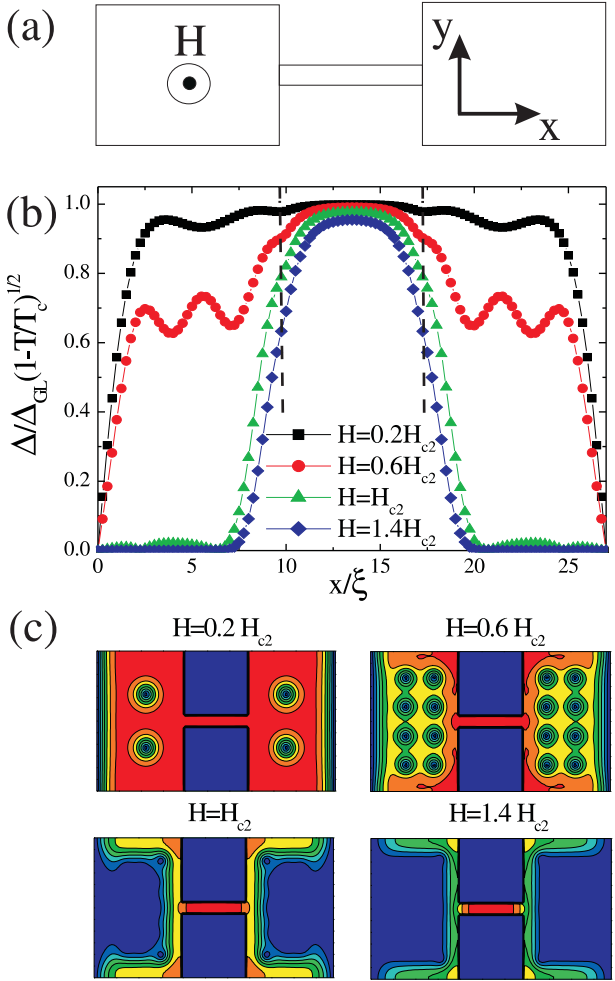}
\caption{(Color online) (a) Schematic illustration of the finite
width Dayem microbridge connected with wide superconducting leads.
(b) Dependence of the magnitude of the order parameter along the
center line of the microbridge for different magnetic fields. (c)
Contour plot of $|\Delta|$ in the leads and microbridge at
different H (blue(red) color corresponds to low(high value for
$|\Delta|$)). The length of the full system (leads+microbridge) is
$27 \xi$, width $15 \xi$, length of microbridge is $7 \xi$ and
width of microbridge is $ \xi$.}
\end{figure}

To study the resistive state of the microbridge we use the set of
equations (time-dependent Ginzburg-Landau equation for the
superconducting order parameter coupled with the kinetic equations
for the quasiparticle distribution function and the Usadel
equations for the Green's functions) derived for 'dirty'
superconductors near the critical temperature of the
superconductor \cite{Schmid1,Larkin,Kramer1,Watts-Tobin}

\begin{subequations}
\begin{eqnarray}
N_1\frac{\partial \delta f_L }{\partial t
}-D\nabla((N_1^2-R_2^2)\nabla \delta f_L) \nonumber
\\
=D\nabla(j_{\epsilon} f_T) -\frac{N_1}{\tau_{in}}\delta
f_L-R_2\frac{\partial f_L^0}{\partial \epsilon}\frac{\partial
|\Delta|}{\partial t},
\end{eqnarray}
\begin{eqnarray}
\frac{\partial}{\partial t}N_1(f_T+e\varphi\frac{\partial
f_L^0}{\partial \epsilon})-D\nabla((N_1^2+N_2^2)\nabla f_T)
\nonumber
\\
=D\nabla(j_{\epsilon}\delta f_L)- \frac{N_1}{\tau_{in}}
\left(f_T+e\varphi\frac{\partial f_L^0}{\partial \epsilon}\right)-
\nonumber
\\
 -N_2|\Delta|\left(2f_T+\hbar\frac{\partial
f_L^0}{\partial \epsilon}\frac{\partial \phi}{\partial t}\right),
\end{eqnarray}
\end{subequations}

\begin{eqnarray}
\hbar D\left(\frac{d^2\Theta}{dx^2}+\frac{d^2\Theta}{dy^2}
\right)+\left(\left(2i\epsilon-\frac{\hbar}{\tau_{in}}\right)-\hbar
DQ^2 \cos\Theta\right)\sin\Theta \nonumber
\\
+2|\Delta|\cos\Theta=0,
\end{eqnarray}

\begin{eqnarray}
\frac{\pi\hbar}{8k_BT_c}\frac{\partial \Delta}{\partial
t}-(\Phi_1+i\Phi_2)\Delta=
\\ \nonumber
\xi_{GL}^2\left(\nabla-i\frac{2eA}{\hbar c}
\right)^2\Delta+\left(1-\frac{T}{T_c}-\frac{|\Delta|^2}{\Delta_{GL}^2}\right)\Delta.
\end{eqnarray}

Here $\Delta=|\Delta|e^{i\phi}$ is a complex order parameter,
$\xi_{GL}^2=\pi\hbar D/8k_BT_c$ and
$\Delta_{GL}^2=8\pi^2(k_BT_c)^2/7\zeta(3)$ are the zero
temperature Ginzburg-Landau coherence length and the order
parameter correspondingly. $Q=(\partial \phi/\partial x-2eA/\hbar
c)$ is a quantity which is proportional to the superfluid velocity
($v_s=DQ$) with A the vector potential taken in the Landau gauge,
$\varphi$ is the electrostatic potential,
$f_L(\epsilon)=f_L^0(\epsilon)+\delta f_L(\epsilon)$ is the
longitudinal and $f_T(\epsilon)$ is the transverse parts of the
quasiparticle distribution function
$2f(\epsilon)=1-f_L(\epsilon)-f_T(\epsilon)$ (in equilibrium
$f_L=f_L^0(\epsilon)=\tanh(\epsilon/2k_BT)$, $f_T^0(\epsilon)=0$).
$N_1$, $N_2$, $R_2$ are the spectral functions which are
determined by the Usadel equation for the normal $\alpha
(\epsilon)=\cos \Theta=N_1(\epsilon)+iR_1(\epsilon)$ and anomalous
$\beta_1 =\beta e^{i\phi}$, $\beta_2=\beta e^{-i\phi}$ ($\beta
(\epsilon)=\sin \Theta=N_2(\epsilon)+iR_2(\epsilon)$) Green
functions.

Nonequilibrium corrections to the quasiparticle distribution
function enters Eq. (3) via the potentials
$\Phi_1=\int_0^{\infty}R_2 \delta f_Ld\epsilon/|\Delta|$ and
$\Phi_2=\int_0^{\infty}N_2f_T d\epsilon/|\Delta|$. Eqs. 1(a,b) are
coupled due to the finite spectral supercurrent
$j_{\epsilon}=Re(\beta_1\nabla \beta_2-\beta_2 \nabla \beta_1
)/2=2N_2R_2Q$.

The advantage of Eqs. (1-3) in comparison with the ordinary or the
extended \cite{Kramer1} time-dependent Ginzburg-Landau equations
is that they allow to take into account nonlocal effects. Here,
under nonlocality we mean that if in one place of the
superconductor and in some moment in time the quasiparticle
distribution function $f(\epsilon)$ becomes nonequilibrium  (due
to some kind of perturbation) then nonequilibrium correction to
$f(\epsilon)$ will be nonzero over a distance $\sim
L_{in}=(D\tau_{in})^{1/2}$ around that point and during a time
$\sim \tau_{in}$ after turning off the perturbation.

Before solving Eqs. (1-3) numerically we calculate the order
parameter in a model 2D system (see Fig. 1) in the stationary
state in the presence of an applied magnetic field. This result
demonstrates the suppression of $|\Delta|$ in the leads by
increasing H and the weak influence on $|\Delta|$ in the
microbridge. It also shows that instead of the 2D model we may use
a 1D model where the suppression of the order parameter in the
superconducting leads could be simulated by introducing locally a
lower critical temperature. Thus we may use the model of Ref.
\cite{Vodolazov2} where we introduce a different critical
temperature in the leads (see Fig. 2). Correlation between $T_c'$
and H is clear: smaller $T_c'$ corresponds to larger H.

\begin{figure}[hbtp]
\includegraphics[width=0.45\textwidth]{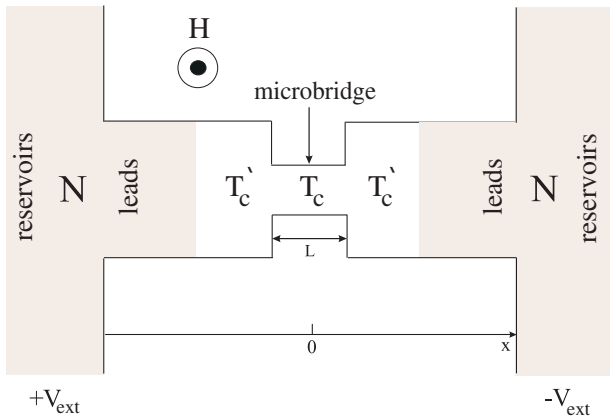}
\caption{Schematic illustration of the model system. Critical
temperature in the leads is smaller than the critical temperature
of the microbridge.}
\end{figure}

The self-consistent set of Eqs. (1-3) was solved numerically using
the method and boundary conditions presented in Ref.
\cite{Vodolazov2}. Further, we use the following dimensionless
units. The order parameter is scaled by $\Delta_0$ ($\Delta_0=1.76
k_BT_c$), distance is in units of the zero temperature coherence
length $\xi_0=\sqrt{\hbar D/\Delta_0}$, time in units of
$t_0=\hbar/\Delta_0$ and temperature in units of the critical
temperature $T_c$. The current is scaled in units of
$j_0=\Delta_0\sigma_n/(\xi_0e)$ and the electrostatic potential is
in units of $\varphi_0=\Delta_0/e$. It is useful to introduce the
dimensionless inelastic relaxation time $\tilde {\tau}_{in}=
\tau_{in}/t_0$ which is the main control parameter in the model
described by Eqs. (1-3). For example in MgB$_2$ $\tilde
{\tau}_{in}\simeq 4$ \cite{Shibata}, in Nb and Pb $\tilde
{\tau}_{in}\simeq 40$, in Sn $\tilde {\tau}_{in}\simeq 70$, in Al
$\tilde {\tau}_{in}\simeq 3500$ and in Zn $\tilde
{\tau}_{in}\simeq 2\cdot 10^4$ \cite{Stuivinga}.

\section{Results}

In Fig. 3 we present the IV characteristics of the superconducting
microbridge with length $L=21 \xi_0$ calculated for different
values of the order parameter in the leads \cite{self1} (in the
inset we show distribution of the time averaged order parameter in
the microbridge and leads at $I=0.75 I_c$). Notice that the
retrapping current {\it increases} and the resistance of the
microbridge {\it decreases} when the order parameter is slightly
suppressed in the leads. We should stress here that the critical
current $I_c$ of the microbridge (at which the superconducting
state becomes unstable) monotonically decreases with decreasing
$T_c'$.
\begin{figure}[hbtp]
\includegraphics[width=0.5\textwidth]{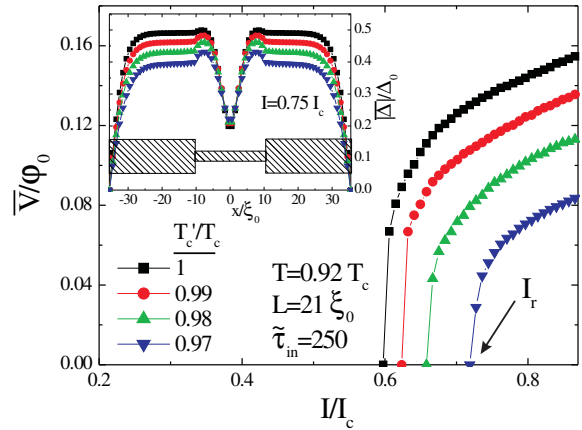}
\caption{(color online). Current voltage characteristics of the
superconducting microbridge with length $L=21 \xi_0\simeq
4.9\xi(T)$ calculated at $T=0.92 T_c$ and $\tilde
{\tau}_{in}=250$. The inset shows the distribution of the
time-averaged order parameter in the microbridge at different
$T_c'$. Current is normalized in units of the critical current of
the microbridge with $T_c'=T_c$.}
\end{figure}

Our result can be explained as due to the enhanced diffusion of
the 'hot' quasiparticles induced by oscillations of the order
parameter in the center of the microbridge \cite{Vodolazov2} when
$\Delta_{lead}$ becomes smaller. Indeed, in this case the energy
barrier connected with the finite energy gap at $\epsilon <
|\Delta|_{lead}$ decreases and nonequilibrium quasiparticles in
the wider energy interval can diffuse to the leads. Here we should
remind the reader that the local energy gap in the microbridge is
smaller than the local value of the order parameter due to the
finite supercurrent and the spatial variation of $|\Delta|$
\cite{Stuivinga,Anthore}. Therefore, 'hot' quasiparticles may
diffuse in the energy interval $\epsilon>|\Delta|_{lead}$ even
when the local $|\Delta|$ in the microbridge is larger than
$|\Delta|_{lead}$ (see inset in Fig. 3).
\begin{figure}[hbtp]
\includegraphics[width=0.52\textwidth]{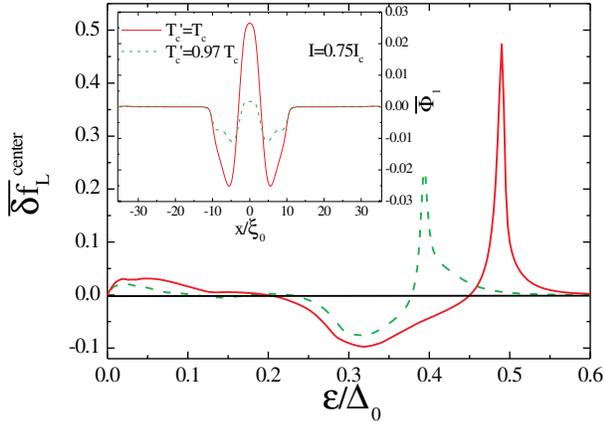}
\caption{(color online). Energy dependence of the time averaged
$\overline{\delta f_L}$ in the center of the microbridge with
length $L=21 \xi_0\simeq 4.9\xi(T)$, $\tilde {\tau}_{in}=250$ and
$T=0.92 T_c$ calculated for $I=0.75 I_c$.}
\end{figure}

In order to illustrate the above effect we show in Fig. 4 the
energy dependence of the nonequilibrium correction to $f_L$ in the
center of the microbridge averaged over one period of Josephson
oscillations. In the inset to Fig. 4 we present the spatial
dependence of the time-averaged potential $\overline{\Phi_1}$
which corresponds to the effective 'temperature'
$T_{eff}=T+T_c\Phi_1$ of quasiparticles in the microbridge
\cite{Vodolazov2}. One can easily see that with decreasing
$\Delta_{lead}$ the energy interval where the quasiparticles are
'heated' (corresponds to a negative sign of $\delta f_L $)
decreases and it results in drastic changes of
$\overline{\Phi_1}$.

The found effect depends strongly on the length of the
microbridge. When $L \lesssim 2\xi(T)$ the value of
$\Delta_{lead}$ has the strongest effect on the value of $I_r$
(which itself is close to the critical current of the microbridge
$I_c$ - see Fig. 5) while the effect of 'heating' is relatively
weak (see Ref. \cite{Vodolazov3}). With decreasing $\Delta_{lead}$
both $I_c$ and $I_r$ monotonically decreases and the resistance
increases - see Fig. 5. The critical length $L_c$ for which $I_r$
starts to increase depends on the inelastic relaxation time - the
larger $\tilde {\tau}_{in}$ the shorter $L_c$. For example $11
\xi_0 <L_c < 15 \xi_0 $ for $\tilde {\tau}_{in}=250$  while $17
\xi_0 <L_c <21 \xi_0 $ for $\tilde {\tau}_{in}=60$ at $T=0.92T_c$.

\begin{figure}[hbtp]
\includegraphics[width=0.45\textwidth]{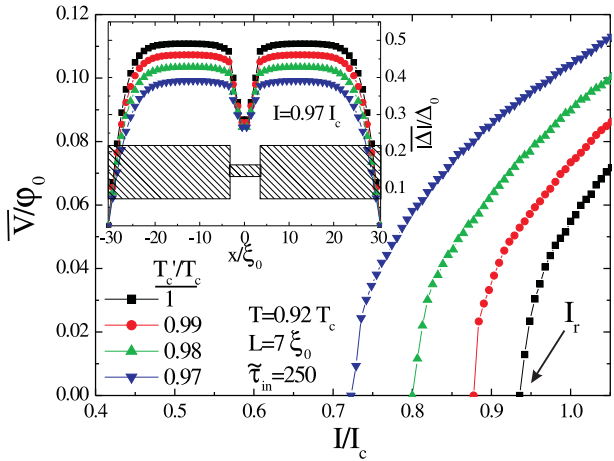}
\caption{(color online). Current voltage characteristics of the
superconducting microbridge with length $L=7 \xi_0\simeq
1.7\xi(T)$ calculated at $T=0.92 T_c$ and $\tilde
{\tau}_{in}=250$. The inset shows the distribution of the
time-averaged order parameter in the microbridge at different
$T_c'$. Current is normalized in units of the critical current of
the microbridge with $T_c'=T_c$.}
\end{figure}

For microbridges with length $L \gg L_{in}$ the relaxation of the
'hot' quasiparticles occurs mainly in the microbridge and hence
the effect of diffusion to the leads is rather weak. This is
illustrated in Fig. 6 where we present IV curves for a relatively
long microbridge $L=31 \xi_0 $ ($\simeq 0.6 L_{in}$ at $\tilde
{\tau}_{in}=250$). The change in the IV curves is much weaker in
comparison with the shorter microbridge (compare with Fig. 3) and
we found that for $L=51 \xi_0 \simeq L_{in}$ the effect
practically disappears for $\tilde {\tau}_{in}=250$.
\begin{figure}[hbtp]
\includegraphics[width=0.52\textwidth]{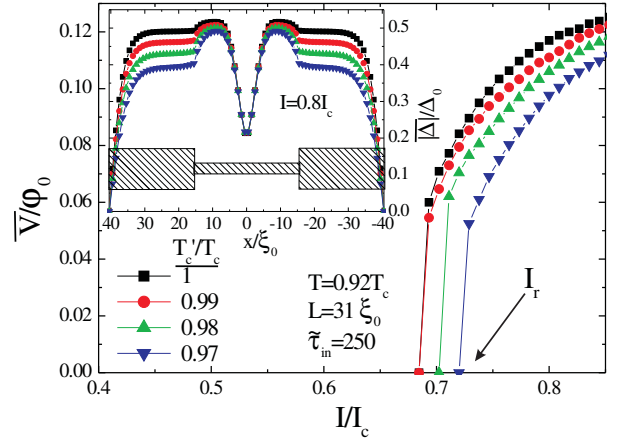}
\caption{(color online). Current voltage characteristics of the
superconducting microbridge with length $L=31 \xi_0 \simeq
7.3\xi(T)$ calculated at $T=0.92 T_c$ and $\tilde
{\tau}_{in}=250$. The inset shows the distribution of the
time-averaged order parameter in the microbridge at different
$T_c'$. Current is normalized in units of the critical current of
the microbridge with $T_c'=T_c$.}
\end{figure}

There is one interesting effect when $L \lesssim L_c$. We find a
decrease of $I_r$ and an increase of the resistance at $I\sim I_r$
but starting from some current $I>I_r$ the resistance of the
microbridge decreases when $\Delta_{lead}$ is suppressed  - see
Fig. 7. At this length there is a competition between the
influence of the order parameter in the leads and the 'heating' of
the quasiparticles on the IV curves. At low currents $I\sim I_r$
the value of $I_r$ is determined mainly by $\Delta_{lead}$ and
$I_r$ decreases with decreasing $\Delta_{leads}$. At larger
currents the 'heating' of the quasiparticles becomes stronger
because it increases with decreasing Josephson period. As a result
decreasing $\Delta_{lead}$ weakens the 'heating' effects and the
voltage at fixed current decreases.
\begin{figure}[hbtp]
\includegraphics[width=0.52\textwidth]{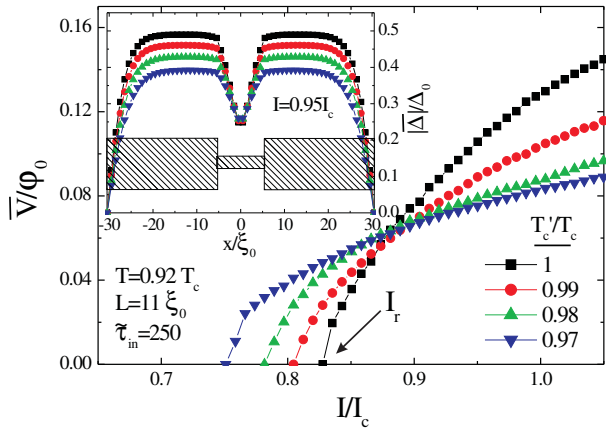}
\caption{(color online). Current voltage characteristics of the
superconducting microbridge with length $L=11 \xi_0 \simeq
2.6\xi(T)$ calculated at $T=0.92 T_c$ and $\tilde
{\tau}_{in}=250$. The inset shows the distribution of the
time-averaged order parameter in the microbridge at different
$T_c'$. Current is normalized in units of the critical current of
the microbridge with $T_c'=T_c$.}
\end{figure}

\section{Discussion}

A typical feature of many experiments about the enhancement of the
critical current and the negative magnetoresistance is presence of
finite resistance of the nanowires/microbridges even at low
temperatures \cite{Tian1,Tian2,Chen1,Chen2,Zgirski} which implies
a strong influence of fluctuations. It can explain the absence of
the hysteresis of IV curves observed in Refs.
\cite{Tian1,Tian2,Chen1,Chen2,Zgirski}. Indeed, it is known for
example from the theory of Josephson junctions that fluctuations
may completely destroy the hysteretic behavior \cite{Tinkham}.
Therefore the current measured in that experiments is probably not
$I_c$ but $I_r$. And indeed, the estimations made in Ref.
\cite{Chen2} showed that the measured critical current was much
smaller (more than 10 times) than the depairing current. Therefore
we believe that our results can be directly applied for the
explanation of the enhancement of the critical current found in
Refs. \cite{Tian1,Tian2,Chen1,Chen2}. Taking into account that in
zinc $L_{in}\simeq 25 \mu m$ \cite{Stuivinga} (when $\xi_0 \simeq
250$ nm \cite{Chen2}) it becomes clear why the effect was weak in
relatively long microbridge with $L=10 \mu m \simeq 0.4 L_{in}$
and in short one with $L=1 \mu m \simeq 4 \xi_0$ \cite{Chen2}. In
Sn the effect was absent for nanowires with $L=6-35 \mu m$
\cite{Tian2} because in tin $\xi_0 \sim 55$ nm and $L_{in} \simeq
470$ nm \cite{Stuivinga}. The applicability of our result to the
experiment of Rogachev et. al \cite{Rogachev} is questionable
because those authors claimed that their critical current is about
the depairing current and furthermore experimental IV curves were
strongly hysteretic (see inset in Fig. 2(c) in Ref.
\cite{Rogachev}). Unfortunately no length dependence of the effect
was studied in that work.

We believe that our result gives a clue in the understanding of
the origin of the negative magnetoresistance at low currents $I
\ll I_r$. It is believed that the finite resistance of the
superconducting nanowires/microbridges at low currents originates
from the finite rate of thermo-activated and/or quantum phase
slips (see for example review \cite{Arutyunov2}). Each phase slip
event is connected with one oscillation of the magnitude of the
order parameter which provides the 'heating' of the
quasiparticles. Therefore, during a finite time
$\min\{\tau_{in},\tau_{diff}\sim L^2/D\}$ after the phase slip
event the effective 'temperature' of quasiparticles will be larger
than the bath temperature and the probability of the next phase
slip becomes higher. It creates the condition for phase slip
avalanches. By decreasing the order parameter in the leads one
increases the flux of 'hot' quasiparticles from the microbridge
and therefore decreases the effective 'temperature' and the
probability of phase slip avalanches. The effect should be
strongest in microbridges with $L \lesssim L_{in}$ where such a
diffusion is the most effective one. In short microbridges with $L
\lesssim 2 \xi(T)$ the suppression of the order parameter in the
leads suppresses also $\Delta$ in the microbridge (due to the
proximity effect - see Figs. 1(b,c)) and it gives the leading
contribution to the probability for phase slips - i.e. it
considerably increases. Therefore one should observe positive
magnetoresistance in short microbridges. In MoGe $L_{in}\simeq 80
nm$ (when $\xi_0\simeq 13 nm$ - see Ref. \cite{Liang}), in Nb
$L_{in} \simeq 170 $nm ($\xi_0$=28 nm \cite{Rogachev}) in Pb
$L_{in} \simeq 240 $nm ($\xi_0$=40 nm), in Al $L_{in} \simeq 8 \mu
m$ ($\xi_0$=140 nm)  and in Zn $L_{in} \simeq 25 \mu m$
($\xi_0$=250 nm). It correlates with the length of the
nanowires/microbridges which were studied in
\cite{Xiong,Rogachev,Tian1,Tian2,Chen1,Chen2,Zgirski} and where
the negative magnetoresistance at low currents were observed both
near and far from $T_c$.

From our theoretical calculations we found that the enhancement of
$I_r$ is absent in the temperature interval $0.92<T/T_c<0.99$ for
$\tilde{\tau}_{in} \lesssim 20$ for all considered lengths of the
microbridges $L=7-51 \xi_0$. This is not surprising because in
this temperature interval the corresponding $L_{in}$ is about the
coherence length and the period of oscillations of the order
parameter is about $\tau_{in}$. For these conditions the effective
'heating' is rather weak (see \cite{Vodolazov3}) and hence
diffusion of 'hot' quasiparticles does not play any role. For
$\tilde{\tau}_{in}=60$ the effect is practically absent at
$T/T_c>0.96$ (in this case $L_{in} \simeq 7.5 \xi_0 \sim \xi(T) $)
and it becomes noticeable at $T/T_c=0.92$ for microbridges with
length $17\xi_0<L<31\xi_0$.

One more interesting effect which comes from our calculations is
the nonmonotonous dependence of $I_r$ on the length of the
microbridge at zero magnetic field (when $T_c'=T_c$). One can see
from comparing the values of $I_r$ in Figs. 3-7 that $I_r$ is
minimal for $L=21 \xi_0$. The reason for such a behavior is the
following - in a very short microbridge the 'heating' of
quasiparticles is weak and $I_r\sim I_c$ and $I_c$ decreases with
decreasing L. In a very long microbridge $L\gg L_{in}$ the 'hot'
quasiparticles relax on the length scale $\sim L_{in}$ near the
phase slip core while in microbridge with length $L<L_{in}$ such a
relaxation is less effective. This results in the following main
tendency: the retrapping current $I_r$ first decreases with
increasing length of the microbridge, reaches the minimal value at
$L \gtrsim 4 \xi$ (when the influence of the leads becomes
sufficiently weak) and than increases and saturates at $L\gg
L_{in}$.

\section{Conclusion}

The 'heating' of quasiparticles, which occurs in the phase slip
core due to oscillations of the order parameter can be reduced by
diffusion of the quasiparticles to the outside regions. It results
in an enhancement of the retrapping current when one slightly
suppresses the order parameter in the superconducting leads. The
enhancement of $I_r$ strongly depends on the length of the
microbridge - the effect is absent in short mictobridges with
length $L\lesssim 2\xi(T)$ and it is weak in relatively long
samples with length $L\gtrsim L_{in}$. Our results predict also a
nonmonotonous dependence of the retrapping current $I_r$ on the
length of the microbridge - it is minimal when $4 \xi(T) \lesssim
L < L_{in}$. We should note that our results cannot be obtained in
the framework of ordinary \cite{Tinkham} or extended
\cite{Kramer1} time-dependent Ginzburg-Landau equations and one
need to solve Eqs. (1-3) where nonlocal effects connected with a
time delay of the response and the diffusion of nonequilibrium
quasiparticles are taken into account.

\begin{acknowledgments}
This work was supported by the Russian Foundation for Basic
Research, Russian Agency of Education under the Federal Target
Programme "Scientific and educational personnel of innovative
Russia in 2009-2013" and the Flemish Science Foundation (FWO-Vl).
\end{acknowledgments}


\begin{references}
\bibitem{self1} We did not present results for smaller values of $T_c'$ because
in this case the order parameter is strongly suppressed near the
ends of the microbridge and the order parameter first starts to
oscillate there. This behavior is out of interest in this paper
because such oscillations create asymmetry in the system with
respect to the center of the microbridge and our numerical method
should be seriously modified and numerically extended to study
such a processes correctly.
\end{references}
\end{document}